\documentstyle[12pt,axodraw]{article}

\newcommand{\be}{\begin{equation}}
\newcommand{\ee}{\end{equation}}
\newcommand{\bea}{\begin{eqnarray}}
\newcommand{\eea}{\end{eqnarray}}

\newcommand{\h}{\hspace{2mm}}

\begin{document}
\baselineskip .25in
\newcommand{\numero}{SHEP 95/24}   %Enter SHEP preprint number

\newcommand{\titre}{One-Loop Quadratic Divergences of the Dual
Non-Linear Sigma Model in Four-Dimensional Spacetime}

\newcommand{\auteura}{Roger D. Simmons}
\newcommand{\place}{Department of Physics\\University of
Southampton\\ Southampton SO17 1BJ \\ U.K. }
\newcommand{\beq}{\begin{equation}}
\newcommand{\eeq}{\end{equation}}

\newcommand{\abstrait}
{Following a review of the dual description of the non-linear sigma
model we investigate the one-loop quadratic divergences. We use the
covariant background field method for the general case and apply the
results to the important example of $SU(2)$}
\begin{titlepage}
\hfill \numero  \\

\vspace{.5in}
\begin{center}
{\large{\bf \titre }}
\bigskip\bigskip \\ by \bigskip \\ \auteura
%\bigskip \\  \auteurb \\
%  \bigskip  and \bigskip \\ \auteurc
\bigskip\bigskip\bigskip \\ \place \bigskip \\

\vspace{.9 in}
{\bf Abstract}
\end{center}
\abstrait
 \bigskip \\
\end{titlepage}

\newpage
\section{Introduction}
Following the recent dramatic work of Seiberg and Witten
\cite{EdAndNatty} in which
a version of Olive-Montonen duality \cite{OlMont}
was found within the context of
$N$=2 supersymmetry, there has been strong motivation to investigate
other possible dual models. One upshot of duality is the inversion of
various coupling terms which may appear in a Lagrangian with the
effect that in some sense the strong and weak coupling regimes of a
theory become interchanged.

The non-linear sigma model has been known for some time to adequately
describe the dynamics of pions \cite{Weinberg} (and mesons in general) at
energies which
are small compared with the inverse confinement radius of QCD.
The non-linear sigma model is inherently versatile and finds
applications to a broad range of physics; in particular it has much to
say about the symmetry breaking sector of the Standard Model and in
this context has been much investigated by Dobado and Herrero
\cite{DobHer}
and Donoghue and Ramirez \cite{RamDon}.
With the promise of interchanging high and
low energy regimes, a study of the dual description of the non-linear
sigma model could be particularly interesting
phenomenologically.

Following the work of
Buscher \cite{BusEtAl} and others \cite{RocPals}
an algorithm exists for the construction of
the dual of the non-linear sigma model in any number of dimensions. The
algorithm consists of gauging a symmetry (an isometry) of the action
by introducing non-propagating gauge fields whose field strength
is constrained to vanish by means of a Lagrange multiplier. Integrating
over the gauge fields instead yields the dual theory where now the
Lagrange multiplier has been promoted to a full dynamical field.

In an earlier paper \cite{MEEEE} we applied this algorithm to a general
four-dimensional non-linear sigma model and found that the Lagrange
multiplier was constrained to be a rank two antisymmetric tensor field,
the dual model being composed of two real scalar fields (the pions
$\pi^\pm$) and the rank two tensor field which was interpreted as the
dual field associated with $\pi^3$. The interesting points of
comparison between the original model and the dual model are expected
to appear in the renormalisation behaviour, and it is this which we
now begin to investigate by looking at the one-loop
quadratic divergences. This should be regarded as nothing more than
just a first step towards a more general study. In particular we shall
have nothing to say about the logarithmic contributions for various
reasons which shall become apparent as we proceed.

We begin by briefly reviewing the dual model.

\section{A Dual Description of the Non-Linear Sigma Model in
Four-Dimensional Spacetime}
We take a target space whose coordinates, $\rho^a$, we split as
$\rho^a=\left(\phi^i ,\theta\right)$ and, without loss of generality,
define a four-dimensional non-linear sigma model on this space as
\be
S\left(\theta ,\phi\right)=
\int d^4x\left[
\frac{1}{2}G\left(\phi\right)\partial^\mu\theta\partial_\mu\theta+
G_i\left(\phi\right)\partial^\mu\phi^i\partial_\mu\theta+
\frac{1}{2}g_{ij}\left(\phi\right)\partial^\mu\phi^i\partial_\mu\phi_j
\right].
\label{Top}
\ee
This action is invariant under the global transformation
$\theta\rightarrow\theta +\alpha$ and the duality emerges upon
minimally gauging this global symmetry and adding a Lagrange
multiplier term constraining the gauge field to be pure gauge
\cite{RocPals}. We
therefore consider
\be
S\left(\theta ,\phi\right)=\int d^4x\left[
\frac{1}{2}G\left(\phi\right)D_\mu\theta D_\mu\theta+
G_i\left(\phi\right)\partial_\mu\phi^iD_\mu\theta+
\frac{1}{2}g_{ij}\left(\phi\right)\partial_\mu\phi^i\partial_\mu\phi^j-
\frac{1}{2}\epsilon_{\mu\nu\rho\sigma}\lambda_{\rho\sigma}F_{\mu\nu}
\right]
\ee
where $D_\mu\theta =\partial_\mu\theta +A_\mu$ and
$F_{\mu\nu}=\partial_\mu A_\nu -\partial_\nu A_\mu$.
$\lambda_{\rho\sigma}$ is the rank two antisymmetric tensor field
referred to in the introduction and appears here as a Lagrange
multiplier enforcing the pure gauge condition. As hinted earlier, the
dual model appears upon integrating out the gauge field and promoting
the Lagrange multiplier to be a fully dynamical gauge field. After
performing this manipulation we obtain \cite{MEEEE}
\be
S\left(\lambda ,\phi\right)=\int d^4x\left[
\frac{1}{2}G_{ij}\partial_\mu\phi^i\partial^\mu\phi^j-
\frac{1}{2G}\epsilon_{\mu\nu\rho\sigma}\partial_\nu\lambda_{\rho\sigma}
\epsilon_{\mu\nu '\rho '\sigma '}\partial_{\nu '}\lambda_{\rho '\sigma
'}+
\frac{1}{G}\epsilon_{\mu\nu\rho\sigma}\partial_\nu\lambda_{\rho\sigma}
G_i\partial_\mu\phi^i\right]
\label{Dual}
\ee
where $G_{ij}$ is given from the tensors appearing in (\ref{Top}) by
\be
G_{ij}=g_{ij}-\frac{1}{G}G_iG_j.
\label{NewG}
\ee

The dual action is therefore seen to be related to (\ref{Top}) via the
interchange
\be
\epsilon_{\mu\nu\rho\sigma}\partial_\nu\lambda_{\rho\sigma}
\rightarrow G_i\partial_\mu\phi^i
\ee
which is in the spirit of the duality encountered in electromagnetism
between the electric and magnetic fields. The apparent problem
associated with the available degrees of freedom (six) of the rank two
antisymmetric field is discussed in detail in \cite{MEEEE} and
\cite{Tow}
where the gauge
freedom is noted to allow just one physical degree of freedom -
consistent with its interpretation as the dual of a real scalar field.

In \cite{MEEEE}
we studied some phenomenology of this model and found that it
reproduced known scattering amplitudes for the charged pions as
derived from the original model. Here we consider the more general
problem of the divergent behaviour of the model.

\section{One-Loop Quadratic Divergences of the Dual Model}
Having arrived at the result (\ref{Dual})
for the dual model with the
explicit inversion of various coupling terms, we now
reparameterise the action such that all coupling terms are upstairs.
This slightly eccentric decision is made purely for aesthetic
reasons to make the following formulae easier to read.

We rewrite the Lagrangian (\ref{Dual}) in the general form
\be
{\cal L}=\frac{1}{2}G_{ij}\partial^\mu\phi^i\partial_\mu\phi^j+
\epsilon_{\mu\nu\rho\sigma}G_i
\partial^\mu\phi^i\partial_\nu A_{\rho\sigma}-\frac{1}{2}
\epsilon_{\mu\nu\rho\sigma}\epsilon_{\mu\nu '\rho '\sigma '}
G\partial_\nu A_{\rho\sigma}\partial_{\nu '}A_{\rho '\sigma '}.
\label{NewDual}
\ee

To investigate the quadratic divergences we first expand the action
using the background field method. This manifestly covariant \cite{Muk1}
procedure is a well known computational tool in quantum field theories
which allows us to compute radiative corrections whilst maintaining
manifestly the symmetries of the theory under consideration
\cite{BigNames}. The
definition of a quantum field which translates as a vector, and the use
of Riemann normal coordinates, leads to an expansion in which each
term depends only on the tensors on the manifold. This geometric
property of the expansion implies a corresponding geometric property
of the counterterms obtained by calculating loop diagrams with the
covariant background field vertices.

Using the background field method we expand the Lagrange density as
\be
{\cal L}[\phi ]=
{\cal L}[\phi_{\mathrm cl}]+
\left. D{\cal L}\right|_{\phi_{\mathrm cl}}+
\frac{1}{2}\left. DD{\cal L}\right|_{\phi_{\mathrm cl}}+...
\ee
with the action of the covariant derivatives being given by \cite{Muk2}
\[
D\partial_\mu\phi^i=\nabla_\mu\xi^i\h\h\h
D\nabla_\mu\xi^i=R^i_{jkl}\partial_\mu\phi^l\xi^j\xi^k
\]
\be
DG=\nabla_iG\xi^i\h\h\h
DG_i=\nabla_jG_i\xi^j\h\h\h
D\nabla_k\nabla_jG_i\xi^k.
\label{CovAct}
\ee
Here $\xi^i$ denote the quantum fluctuations of the $\phi^i$ fields
whilst for the rank two antisymmetric tensor field, $A_{\rho\sigma}$,
the quantum fluctuations are denoted $\lambda_{\rho\sigma}$ and the
action of the covariant derivatives is
\be
D\partial_\nu A_{\rho\sigma}=\partial_\nu\lambda_{\rho\sigma}
\ee

Upon expanding the Lagrange density (\ref{NewDual})
and inserting the expressions
(\ref{CovAct}) where appropriate,
we arrive at a propagator term for the $\xi^i$ fields of the form
\be
G_{ij}\nabla_\mu\xi^i\nabla_\mu\xi^j.
\ee
This can be manipulated into a useable form by moving to the tangent
space via the following definitions \cite{Egu}
\be
\xi^a=e^a_i\xi^i\h\h\h \xi^i=E^i_a\xi^a,
\ee
allowing us to write
\be
G_{ij}\nabla_\mu\xi^i\nabla_\mu\xi^j\rightarrow
e^a_ie^b_j\eta_{ab}\nabla_\mu\xi^i\nabla_\mu\xi^j =
\eta_{ab}\nabla_\mu\xi^a\nabla_\mu\xi^b
\ee
where use has been made of the relations $G_{ij}=e^a_ie^b_j\eta_{ab}$
and $\nabla_\mu e^a_i=0$. Having made the shift to the tangent space
our covariant derivative is now defined via the spin connection,
$\omega^a_{ib}$,
\be
\nabla_\mu\xi^a =
\partial_\mu\xi^a +\omega^a_{ib}\xi^b\partial_\mu\phi^i.
\ee
Applying this transformation to all $\xi^i$ terms we parameterise
the Lagrangian for the quantum fields as
\begin{eqnarray}
{\cal L}=\frac{1}{2}DDS &=&
\frac{1}{2}\eta_{ab}\partial^\mu\xi^a\partial_\mu\xi^b-\frac{1}{2}
\epsilon_{\mu\nu\rho\sigma}\epsilon_{\mu\nu '\rho '\sigma '}
\partial_\nu\widetilde{\lambda}_{\rho\sigma}
\partial_{\nu '}\widetilde{\lambda}_{\rho '\sigma '}\nonumber\\
\ &+& A_{ab}\xi^a\xi^b +B^\mu_{ab}\xi^a\partial_\mu\xi^b
+C^{\mu\nu\rho}_a\xi^a\partial_\mu\widetilde{\lambda}_{\nu\rho}+
D^{\mu\nu}_a\xi^a\widetilde{\lambda}_{\mu\nu}\nonumber\\
\ &+&
E^{\mu\rho\sigma}_a\partial_\mu\xi^a\widetilde{\lambda}_{\rho\sigma}
+F^{\mu\nu\rho\sigma}\widetilde{\lambda}_{\mu\nu}
\widetilde{\lambda}_{\rho\sigma}+
G^{\mu\nu\rho\sigma\tau}\widetilde{\lambda}_{\mu\nu}\partial_\rho
\widetilde{\lambda}_{\sigma\tau}
\end{eqnarray}
where, to get the kinetic term for the $\lambda_{\rho\sigma}$ fields
into a convenient form, we have introduced
$\widetilde{\lambda}_{\rho\sigma}=\sqrt{G}\lambda_{\rho\sigma}$ with
the suitably covariantised derivative
\be
\partial_\nu\lambda_{\rho\sigma}\rightarrow
\frac{1}{\sqrt{G}}\partial_\nu\widetilde{\lambda}_{\rho\sigma}-
\frac{1}{2}\frac{1}{G\sqrt{G}}
\nabla_mG\partial_\nu\phi^m\widetilde{\lambda}_{\rho\sigma}.
\ee

Using this parameterisation we obtain
\begin{eqnarray}
2A_{ab}&=&
\partial^\mu\phi^i\partial_\mu\phi^j\left(
\eta_{cd}\omega^c_{ia}\omega^d_{jb}+R_{jmki}E^m_aE^k_b\right)
\nonumber\\
&+&\epsilon_{\mu\nu\rho\sigma}\partial_\nu A_{\rho\sigma}\left(
\nabla_k\nabla_jG_iE^j_aE^k_b+G_mR^m_{jki}E^j_aE^k_b+
2\nabla_jG_m\omega^c_{ib}E^m_cE^j_a\right)\nonumber\\
&-&\frac{1}{2}\epsilon_{\mu\nu\rho\sigma}
\epsilon_{\mu\nu '\rho '\sigma '}\partial_\nu A_{\rho\sigma}
\partial_{\nu '}A_{\rho '\sigma '}\nabla_j\nabla_iGE^i_aE^j_b
\nonumber\\
\ & \ & \ \nonumber\\
B^\mu_{ac}&=&\left(
\eta_{cb}\omega^b_{ia}\partial_\mu\phi^i +
\epsilon_{\mu\nu\rho\sigma}\partial_\nu A_{\rho\sigma}
\nabla_jG_iE^i_cE^j_a\right)\nonumber\\
\ & \ & \ \nonumber\\
C^{\nu\rho\sigma}_a&=&
\frac{1}{\sqrt{G}}\epsilon_{\mu\nu\rho\sigma}\left[\left(
2\nabla_{[j}G_{i]}+
\frac{1}{2G}G_j\nabla_iG\right)E^j_a\partial_\mu\phi^i-
\epsilon_{\mu\nu '\rho '\sigma '}\partial_{\nu '}A_{\rho '\sigma '}
\nabla_iGE^i_a\right]\nonumber\\
\ & \ & \ \nonumber\\
D^{\rho\sigma}_a&=&
-\frac{\nabla_jG\partial_\nu\phi^j}{2G\sqrt{G}}
\epsilon_{\mu\nu\rho\sigma}
\left[\left(
\nabla_mG_iE^m_a+G_m\omega^d_{ia}E^m_d\right)\partial_\mu\phi^i-
\epsilon_{\mu\nu '\rho '\sigma '}\partial_{\nu '}A_{\rho '\sigma '}
\nabla_iGE^i_a\right]\nonumber\\
\ & \ & \ \nonumber\\
E^{\mu\rho\sigma}_a&=&
-\frac{1}{2}\frac{1}{G\sqrt{G}}\epsilon_{\mu\nu\rho\sigma}
\partial_\nu\phi^iG_j\nabla_iGE^j_a\nonumber\\
\ & \ & \ \nonumber\\
F^{\rho\sigma\rho '\sigma '}&=&
-\frac{1}{8G^2}
\epsilon_{\mu\nu\rho\sigma}\epsilon_{\mu\nu '\rho '\sigma '}
\partial_\nu\phi^i\partial_{\nu '}\phi^j\nabla_iG\nabla_jG
\nonumber\\
\ & \ & \ \nonumber\\
G^{\rho '\sigma '\nu\rho\sigma}&=&
\frac{1}{2G}\epsilon_{\mu\nu\rho\sigma}
\epsilon_{\mu\nu '\rho '\sigma '}\partial_{\nu '}\phi^i
\nabla_iG\h .
\end{eqnarray}

Using cut-off regularisation at the scale $\Lambda$, the Feynman
diagrams which contribute $\sim\Lambda^2$ divergences are
(dashed lines = $\xi$, coiled lines = $\lambda$):
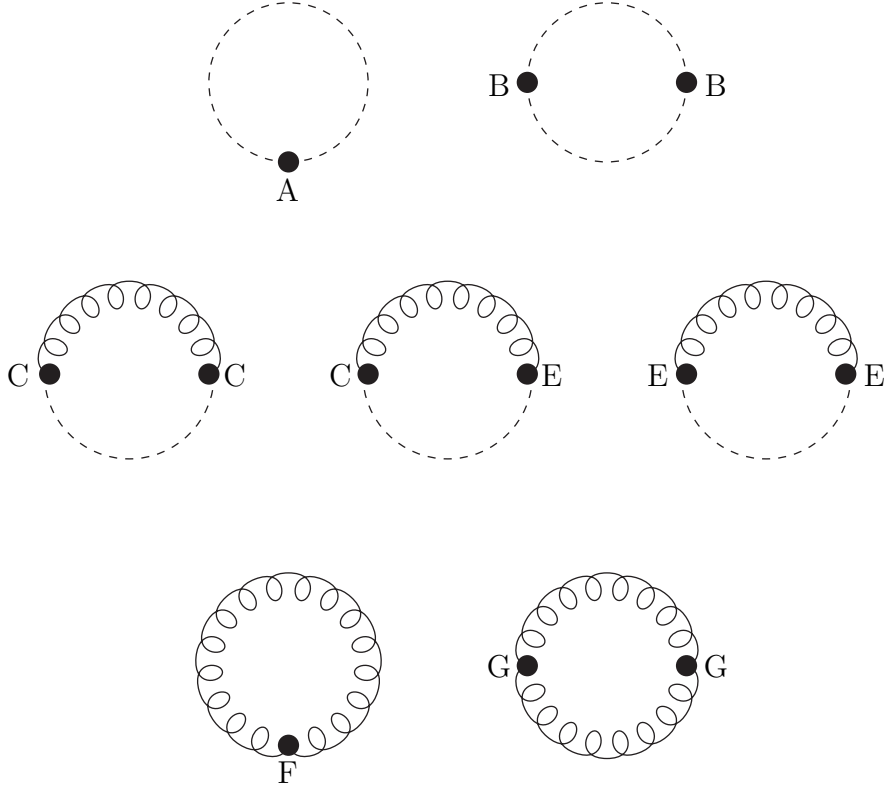
\begin{center}
\begin{picture}(300,300)(0,0)

%% Bottom two

\GlueArc(90,40)(30,-90,270){5}{16}
\Vertex(90,10){4}
\Text(90,0)[]{F}

\GlueArc(210,40)(30,0,180){5}{8}
\GlueArc(210,40)(30,180,360){5}{8}
\Vertex(180,40){4}
\Vertex(240,40){4}
\Text(170,40)[]{G}
\Text(252,40)[]{G}

%% Middle three

\DashCArc(30,150)(32,180,360){3}
\GlueArc(30,150)(30,0,180){5}{8}
\Vertex(0,150){4}
\Vertex(60,150){4}
\Text(-12,150)[]{C}
\Text(70,150)[]{C}

\DashCArc(150,150)(32,180,360){3}
\GlueArc(150,150)(30,0,180){5}{8}
\Vertex(120,150){4}
\Vertex(180,150){4}
\Text(110,150)[]{C}
\Text(190,150)[]{E}

\DashCArc(270,150)(32,180,360){3}
\GlueArc(270,150)(30,0,180){5}{8}
\Vertex(240,150){4}
\Vertex(300,150){4}
\Text(230,150)[]{E}
\Text(312,150)[]{E}

%% Top two

\DashCArc(90,260)(30,-90,270){3}
\Vertex(90,230){4}
\Text(90,220)[]{A}

\DashCArc(210,260)(30,0,180){3}
\DashCArc(210,260)(30,180,360){3}
\Vertex(180,260){4}
\Vertex(240,260){4}
\Text(170,260)[]{B}
\Text(252,260)[]{B}

\end{picture}
\end{center}
Evaluating these diagrams we finally obtain
the one-loop effective Lagrangian
\be
{\cal L}_{{\mathrm eff}}=
\frac{1}{2}\partial^\mu\phi^i\partial_\mu\phi^jG^R_{ij}+
\epsilon_{\mu\nu\rho\sigma}\partial_\nu A_{\rho\sigma}
\partial_\mu\phi^iG^R_i-
\frac{1}{2}\epsilon_{\mu\nu\rho\sigma}
\epsilon_{\mu\nu '\rho '\sigma '}\partial_\nu A_{\rho\sigma}
\partial_{\nu '}A_{\rho '\sigma '}G^R
\label{618}
\ee
with the renormalised tensors
\begin{eqnarray}
G^R_{ij} &=& G_{ij}+J\left[
R_{ij}-\frac{6}{G}G^{mn}\nabla_{[n}G_{i]}\nabla_{[m}G_{j]}\right]
\nonumber\\
\ & \ & \ \nonumber\\
G^R_i &=& G_i-\frac{J}{2}\left[
G^{mn}\nabla_n\nabla_mG_i-G^jR_{ji}-
\frac{3}{G}G^{mn}\nabla_mG\nabla_{[n}G_{i]}\right]\nonumber\\
\ & \ & \ \nonumber\\
G^R &=& G+\frac{J}{2}\left[
G^{mn}\left(
\frac{3}{G}\nabla_mG-\nabla_m\right)
\nabla_nG
-\nabla_jG_i\nabla_mG_n\left(
G^{mj}G^{in}-G^{jn}G^{im}\right)\right]\nonumber\\
\ & \ & \
\label{620}
\end{eqnarray}
and
\be
J=\int^\Lambda_0 d^4k\frac{1}{k^2}=-\frac{1}{16\pi^2}\Lambda^2
\ee
where we have Wick rotated and
$d^4k=\frac{1}{2}k^2\sin^2\theta\sin\phi dk^2d\theta d\psi$ is the
integration measure in Euclidean space.

This is our main result and represents the general one-loop
corrected effective Lagrangian for the Abelian dualed sigma model.
\footnote{We are only interested here in divergences $\sim\Lambda^2$
and ignore the logarithmic contributions which can only be sensibly
interpreted in the full momentum expansion of which the sigma
model just described is the $O(p^2)$ leading term.}
Having derived the dual version of the non-linear sigma model and
the very general renormalisation to one-loop, we now specialise
to the important case of $SU(2)$.

\section{Application to $SU(2)$}
We take as our starting point the non-linear
sigma model parameterised by the matrix field $U(x)$ belonging
to the quotient space $SU(2)_L\otimes SU(2)_R/SU(2)_{L+R}$,
\be
U(x)=\exp\left(i\tau .\phi/\Lambda\right),
\ee
where $\phi^a$, $a=1,2,3$, are the Goldstone bosons associated with
the symmetry breaking
$SU(2)_L\otimes SU(2)_R\rightarrow SU(2)_{L+R}$,
$\tau^a$ are the $2\times 2$ Pauli matrices and $\Lambda$ some
energy scale. (When applied to the symmetry breaking sector of the
Standard Model $\Lambda$ is fixed at the scale of the Higgs VEV =
246GeV).

Under an $SU(2)_L\otimes SU(2)_R$ transformation the matrix $U$
transforms as $U\rightarrow LUR^\dagger$ and the $SU(2)$ invariant
sigma model can be written
\be
{\cal L}=
\frac{\Lambda^2}{4}{\mathrm Tr}\partial^\mu U\partial_\mu U^{-1}.
\label{612}
\ee
To obtain the dual version of this model we must first massage
(\ref{612}) into the form (\ref{Top})
which we achieve by separating out one field -
the field which shall carry the Abelian symmetry. When this is done
(see \cite{MEEEE} for definitions)
we recover the form of (\ref{Top}) with
%Choosing to
%isolate the field $(\theta)$ associated with $\tau^3$ we have
%\begin{eqnarray}
%{\cal L}&=&
%\frac{1}{2}\partial^\mu\theta\partial_\mu\theta+
%\frac{1}{2}\partial^\mu\pi\partial_\mu\pi\frac{\sin^2\Omega}{\Omega^2}
%+\frac{1}{\Lambda}\epsilon_{ij}\pi^j\partial^\mu\theta\partial_\mu\pi^i
%\frac{\sin^2\Omega}{\Omega^2}\nonumber\\
%&+& \frac{1}{2\Lambda^2}\frac{1}{\Omega^2}\left[
%1-\frac{\sin^2\Omega}{\Omega^2}\right]\left(
%\pi .\partial^\mu\pi\right)^2
%\end{eqnarray}
%thereby recovering the form of (??)with
\begin{eqnarray}
g_{ij}&=&
\delta^{ij}\frac{\sin^2\Omega}{\Omega^2}+
\frac{1}{\Omega^2\Lambda^2}\left[
1-\frac{\sin^2\Omega}{\Omega^2}\right]\pi^i\pi^j\nonumber\\
G_i&=&
\epsilon_{ij}\frac{1}{\Lambda}\pi^j\frac{\sin^2\Omega}{\Omega^2}
\h\h\h G=1.
\label{616}
\end{eqnarray}
Once in this form we use the rules given in (\ref{Dual}) and
(\ref{NewG}) to write
down the dual verison of the $SU(2)$ non-linear sigma model to $O(p^2)$
%\footnote{For this purely formal discussion we stop at $O(p^2)$ terms.
%The only really sensible interpretation of (??) is as the first term
%in a general momentum expansion which is not relevant to our discussion of
%the quadratic divergences, and which we therefore ignore.}
\begin{eqnarray}
{\cal L}&=&
-\frac{1}{2}\epsilon_{\mu\nu\rho\sigma}\partial_{\nu}\lambda_{\rho\sigma}
\epsilon_{\mu\nu '\rho '\sigma '}\partial_{\nu'}\lambda_{\rho'\sigma'}
+\frac{1}{\Lambda}\epsilon_{\mu\nu\rho\sigma}\partial_\nu
\lambda_{\rho\sigma}\epsilon_{ij}\pi^j\partial_\mu\pi^i\frac
{\sin^2\Omega}{\Omega^2}\nonumber\\
&+&\frac{1}{2}\partial^\mu\pi^i\partial_\mu\pi^i
\frac{\sin^2\Omega}{\Omega^2}\left(
1-\sin^2\Omega\right)+
\frac{1}{2}\frac{1}{\Lambda^2\Omega^2}\left[
1-\frac{\sin^2\Omega}{\Omega^2}+\frac{\sin^4\Omega}{\Omega^2}\right]
\left(\pi .\partial_\mu\pi\right)^2,\nonumber\\
\label{621}
\end{eqnarray}
where in the language of (\ref{NewDual}) we have
\[
G_{ij}=
\delta^{ij}\frac{\sin^2\Omega}{\Omega^2}\left[
1-\sin^2\Omega\right]+
\frac{\pi^i\pi^j}{\Lambda^2\Omega^2}\left[
1-\frac{\sin^2\Omega}{\Omega^2}+\frac{\sin^4\Omega}{\Omega^2}\right]
\]
\be
G_i=\epsilon_{ij}\frac{1}{\Lambda}\pi^j\frac{\sin^2\Omega}{\Omega^2}
\h\h\h G=1.
\label{619}
\ee

We can now immediately use (\ref{618}) to investigate the
$\sim\Lambda^2$ divergences appearing at the one-loop level in the
dual $SU(2)$ non-linear sigma model. Given the relations in
(\ref{619}) and
inserting into (\ref{620})
we arrive at the surprisingly compact expression
\begin{eqnarray}
{\cal L}&=&
\frac{1}{2}\partial^\mu\pi^i\partial_\mu\pi^jG_{ij}(1-2J)-
\frac{1}{2}\epsilon_{\mu\nu\rho\sigma}\epsilon_{\mu\nu '\rho '\sigma'}
\partial_\nu\lambda_{\rho\sigma}\partial_{\nu '}\lambda_{\rho '\sigma '}
(1-2J)\nonumber\\
&+&\epsilon_{\mu\nu\rho\sigma}\partial_\nu\lambda_{\rho\sigma}
\partial_\mu\pi^iG_i.
\end{eqnarray}
This is the one-loop corrected effective $SU(2)$ Lagrangian for the
Abelian dualed sigma model and can be compared with the results
presented in \cite{MKG}
for the conventional non-linear sigma model.

\section{Conclusions}
We have presented a completely general discussion of the quadratic
renormalisation behaviour of the dual non-linear sigma model at the
one loop level and applied it to the important phenomenological case
of $SU(2)$. This is the first step in a general study of the divergent
behaviour of the dual non-linear sigma model where we hope to find a
formalism which will facilitate a study of the infrared behaviour
(fixed points) which in itself will enable a detailed
comparison between this
model and the standard non-linear sigma model.

The next stage in the programme will be to extend the investigations
to accommodate the logarithmic divergences. Here there is a
considerable increase in the number of Feynman diagrams which we need
to consider to one loop ($\approx 20$) and in addition the infrared
singularities which emerge as the rank two antisymmetric tensor fields
go soft in certain diagrams demand careful attention. It is hoped to
pursue this line of investigation in a subsequent paper.

\begin{center}
{\bf Acknowledgements}
\end{center}
The author gratefully acknowledges N. Mohammedi for his guidance
throughout
the duration of this study, and R.T. Moss who checked some of the
results presented here. This work was supported in part by PPARC.

\end{document}